\documentclass[twocolumn]{aastex63}

\usepackage{natbib}

\received{June 24, 2020}
\revised{July 24, 2020}
\accepted{July 24, 2020}
\submitjournal{ApJ}

\begin{document}

\title{Dynamics of Late-Stage Reconnection in the 2017 September 10 Solar Flare}

\author[0000-0001-9726-0738]{Ryan J. French}
\affil{Mullard Space Science Laboratory, University College London, Dorking, RH5 6NT, UK}

\author[0000-0001-9346-8179]{Sarah A. Matthews}
\affil{Mullard Space Science Laboratory, University College London, Dorking, RH5 6NT, UK}

\author[0000-0002-2943-5978]{Lidia van Driel-Gesztelyi}
\affil{Mullard Space Science Laboratory, University College London, Dorking, RH5 6NT, UK}\affil{LESIA, Observatoire de Paris, Université PSL, CNRS, Sorbonne Université, Université Paris Diderot, 5 place Jules Janssen, 92190 Meudon, France}
\affil{Konkoly Observatory of the Hungarian Academy of Sciences, Budapest, Hungary}

\author[0000-0003-3137-0277]{David M.~Long}
\affil{Mullard Space Science Laboratory, University College London, Dorking, RH5 6NT, UK}

\author[0000-0001-5174-0568]{Philip G. Judge}
\affil{HAO, National Center for Atmospheric Research, P.O. Box 3000, Boulder CO 80307-3000, USA}

\begin{abstract}

In this multi-instrument paper, we search for evidence of sustained magnetic reconnection far beyond the impulsive phase of the X8.2-class solar flare on 2017 September 10. Using Hinode/EIS, CoMP, SDO/AIA, K-Cor, Hinode/XRT, RHESSI, and IRIS, we study the late-stage evolution of the flare dynamics and topology, comparing signatures of reconnection with those expected from the standard solar flare model. 
Examining previously unpublished EIS data, we present the evolution of non-thermal velocity and temperature within the famous plasma sheet structure, for the first four hours of the flare's duration. 
On even longer time scales, we use Differential Emission Measures and polarization data to study the longevity of the flare's plasma sheet and cusp structure, discovering that the plasma sheet is still visible in CoMP linear polarization observations on 2017 September 11, long after its last appearance in EUV. We deduce that magnetic reconnection of some form is still ongoing at this time -- 27 hours after flare onset.

\end{abstract}


\section{Introduction} 

Solar flares are a key form of energy release in the corona, as free energy stored in coronal magnetic fields converts into kinetic, thermal, and electromagnetic energy through the process of magnetic reconnection. Powered by the inflow of plasma with oppositely orientated magnetic field components, magnetic reconnection is in essence a dynamic change in magnetic topology to a lower energy configuration. 
In the standard “CSHKP” solar flare model
\citep{Carmichael,Sturrock,Hirayama,Kopp}, reconnection is believed to power the impulsive phases of flares, typically lasting a few minutes to an hour.
There has, however, been evidence to suggest that magnetic reconnection within flares can persist for much longer than this, with energy release timescales of tens of hours \citep{Bruzek64}, albeit at a more gradual rate than during the impulsive phase. 

2017 September 10 saw the eruption of an X8.2-class flare on the western solar limb. 
Multiple aspects of this flare have been studied, in particular the long-lived plasma sheet structure, which rose above the solar limb behind the erupting coronal mass ejection (CME), where it remained for several hours. 
Due to high temperatures and enhanced line widths in the region \citep{Warren}, the plasma sheet has been interpreted as hot, turbulent plasma, enclosing the current sheet at the site of magnetic reconnection. 
Due to the unique edge-on perspective of the eruption, and its visually striking similarities to the standard “CSHKP” 2D solar flare model, this particular plasma sheet has prompted many studies of the nature of reconnection and flare energy releases. 
Additionally, \citet{Chen20} found that the 2017 September 10 event fits the 3D models of eruptive flares \citep{Aulanier2012,Aulanier2013,Janvier2013,Janvier}, finding that the plasma sheet is sitting above the underlying arcade along (and arching over) the east-west inversion line in the line-of-sight (LOS). The plasma sheet extends to lower altitudes in the foreground, and is briefly visible above the continuation of the arcade in the north-south direction (at the footpoints of the erupting flux rope), to the south of the primary flare arcade \citep{Cai}.

Observations of the impulsive phase of the flare have been thoroughly explored. Multiple
studies have revealed evidence for turbulence or tearing mode instabilities within the plasma sheet, such as 
unusually high 
emission line widths, and associated non-thermal velocities, which \citet{Warren} interpreted as reconnection-induced turbulence and outflows. Quasi-periodic pulsations have been 
reported \citep{Longcope,Cheng,Hayes}, along with small-scale velocity fluctuations consistent with turbulent plasmoid fragmentation \citep{Cheng}. Most of these observations occurred within forty minutes of flare onset. Reconnection outflows along the plasma sheet during the post-impulsive phase of this flare are explored in \citet{Yu2020}.

\citet{French} provided further evidence to suggest the presence of reconnection-induced instabilities, based upon 
Coronal Multi-channel Polarimeter (CoMP) spectropolarimetric data from coronal forbidden Fe XIII lines. In spite of the strikingly laminar
appearance of the plasma sheet in unpolarized measurements (e.g. from AIA),
the amplitude of linearly polarized light was found to be anomalously small 
along the sheet. In a process of elimination, 
the authors concluded that the magnetic field could not be laminar, and instead must contain unresolved structure on sub-pixel scales. 
Given theoretical 
work on plasmoid formation during reconnection, these data were interpreted by \citet{French} as evidence for the presence of instabilities within the reconnecting plasma sheet. 
These CoMP measurements took place 4 hours after the flare onset, long beyond the impulsive phase of the event. The plasma sheet is still clearly visible in extreme ultraviolet (EUV) images at this time, \citep[similar to the duration of previously observed EUV plasma sheets,][]{Savage2010,Seaton2017}.
The presence of sustained reconnection and instabilities within the plasma sheet at this point in the flare's evolution suggests that the reconnection process plays a role throughout the entire dynamic evolution of the flare, and not just during the impulsive phase.

In this study, we provide additional evidence for sustained fast reconnection late in the flare's evolution, by examining plasma kinematics associated with the standard flare model, beyond the impulsive phase of the event. These features include Doppler velocities, hard X-ray (HXR) sources, and flare loop growth. We also track the evolution of non-thermal velocity at the base of the plasma sheet, supporting the interpretation from the 2017 September 10 CoMP data that the low polarization signature is likely a result of internal magnetic structure expected from the presence of small-scale plasma instabilities.
On even longer timescales, we examine 2017 September 11 CoMP observations, showing that magnetic reconnection is still occurring in the plasma sheet structure, over a day into the event (and several hours after its last appearance in EUV). 

\section{Observations}
\begin{figure}
  \centering
  \includegraphics[width=8.5cm]{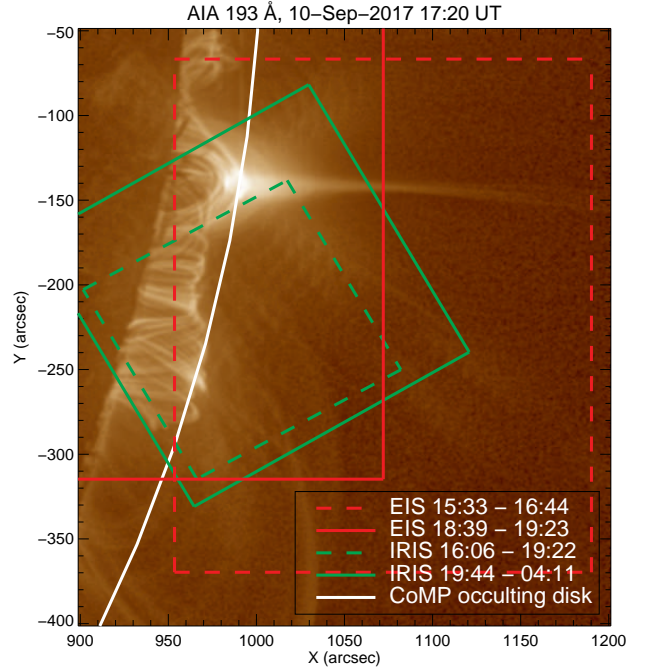}
  \caption{IRIS (green) and EIS (red) FOVs overlaid on an AIA 193 {\AA} image. Dashed lines represent earlier observing sequences, and solid lines later. The location of the CoMP occulting disk is shown in white.}
  \label{fig:FOV}
\end{figure}

The 2017 September 10 event was observed across the electromagnetic spectrum by multiple instruments. 
The Hinode Extreme Ultraviolet Imaging Spectrometer \citep[EIS;][]{Culhane} captured the flare from its onset at 15:44 UT, past its peak at 16:06 UT, up to 16:52 UT. 
Observations later continued at 18:39 UT, observing the late evolution of the flare until 19:31 UT. 
In both observing runs, EIS observed the flare with a $2 \arcsec$ slit, which provides an approximately $4 \arcsec$ resolution over a field of view (FOV) of $240 \times 300 \arcsec$. 
The slit spectrometer rasters the FOV from west to east with a cadence of 8 minutes 52 seconds. 
The plasma sheet is most visible in the Fe XXIV 192.04 {\AA} line, which samples plasma with a typical formation temperature of 17 MK in a thermal plasma. 
The FOV changed between the two EIS observation sequences. The first sequence observed the full radial extent of the plasma sheet, whereas the second was pointed $120 \arcsec$ to the east, missing the upper plasma sheet but sampling some of the underlying the solar disk. 
Both FOVs are shown in Figure \ref{fig:FOV}.

Spectral maps from EIS are supported by full disk EUV images from the Solar Dynamics Observatory (SDO) Atmospheric Imaging Assembly \citep[AIA;][]{Lemen}. AIA observed the event with
a 12 second cadence and $0.6 \arcsec$ spatial resolution, with a varying exposure time. The plasma sheet is most visible in the broadband 193 {\AA} filter, which measures plasma from the cooler Fe XII and hotter Fe XXIV lines. It is emission from this hotter line dominating during the early flare, which explains the similarities between AIA 193 {\AA} and EIS 192.04 {\AA} observations (Figure \ref{fig:eis}). 

The Hinode X-Ray Telescope \citep[XRT;][]{Golub} captured the duration of the flare in thin Al Poly and thin Be filters. This is a broad band imaging instrument, with temperature sensitivity peaking at 6.9 and 7 MK respectively for each filter \citep{Narukage2011}. Observations continued for the days succeeding the flare and imaged the post-flare loops as they rotated off the limb.

Observations from the Interface Region Imaging Spectrograph \citep[IRIS;][]{DePontieu} are also available for this event. IRIS observed from 12:00 UT to 19:22 UT with a \textit{large} ($119 \times 119 \arcsec$) slit jaw FOV orientated at 60 deg, capturing the edge of the erupting flare in the top left of the frame. At 19:44, IRIS slit jaw observations switched to a \textit{very large} ($166 \times 175 \arcsec$) FOV, continuing until 04:13 UT. In this wider FOV, we observe the larger flare loop structure and primary plasma sheet base. In both the \textit{large} and \textit{very large} FOV rasters, IRIS acquired data with the 1330 {\AA} slit-jaw window. The FOV of both rasters are shown in Figure \ref{fig:FOV}.

These observations were complemented by HXR measurements from the Reuven Ramaty High Energy Solar Spectroscopic Imager \citep[RHESSI;][]{Lin}. RHESSI had four operational detectors at the time of the flare, and took measurements intermittently across the flare's duration. 

Finally, we include observations from the Mauna Loa Solar Observatory (MLSO) Coronal Multi-channel Polarimeter (CoMP) \citep{CoMP} and K-coronagraph (K-Cor) instruments. 2017 September 10 data from these instruments have been presented in \citet{French}. Mauna Loa observations continued the next local morning on 2017 September 11, from 17:41 and 17:37 UT for CoMP and K-Cor respectively. CoMP measures linear spectropolarimetry (Stokes I,Q,U) of infrared coronal forbidden lines in the low corona ($\sim$1.03 to 1.5 \(R_\odot\)), with the occulting disk location shown in Figure \ref{fig:FOV}. Here we focus on Fe XIII 1074.7 nm emission, measured with a $4.35 \arcsec$ spatial resolution and a varying cadence. Fe XIII 1079.4 nm are also available, forming a density sensitive pair with Fe XIII 1074.7 nm. K-Cor complements this with measurements of white light polarization (pB) from $\sim$1.03 to 1.5 \(R_\odot\), with a $5.64 \arcsec$ spatial sampling and 15s cadence.

\section{EIS Spectroscopy}

\begin{figure*}
  \centering
  \includegraphics[width=16cm]{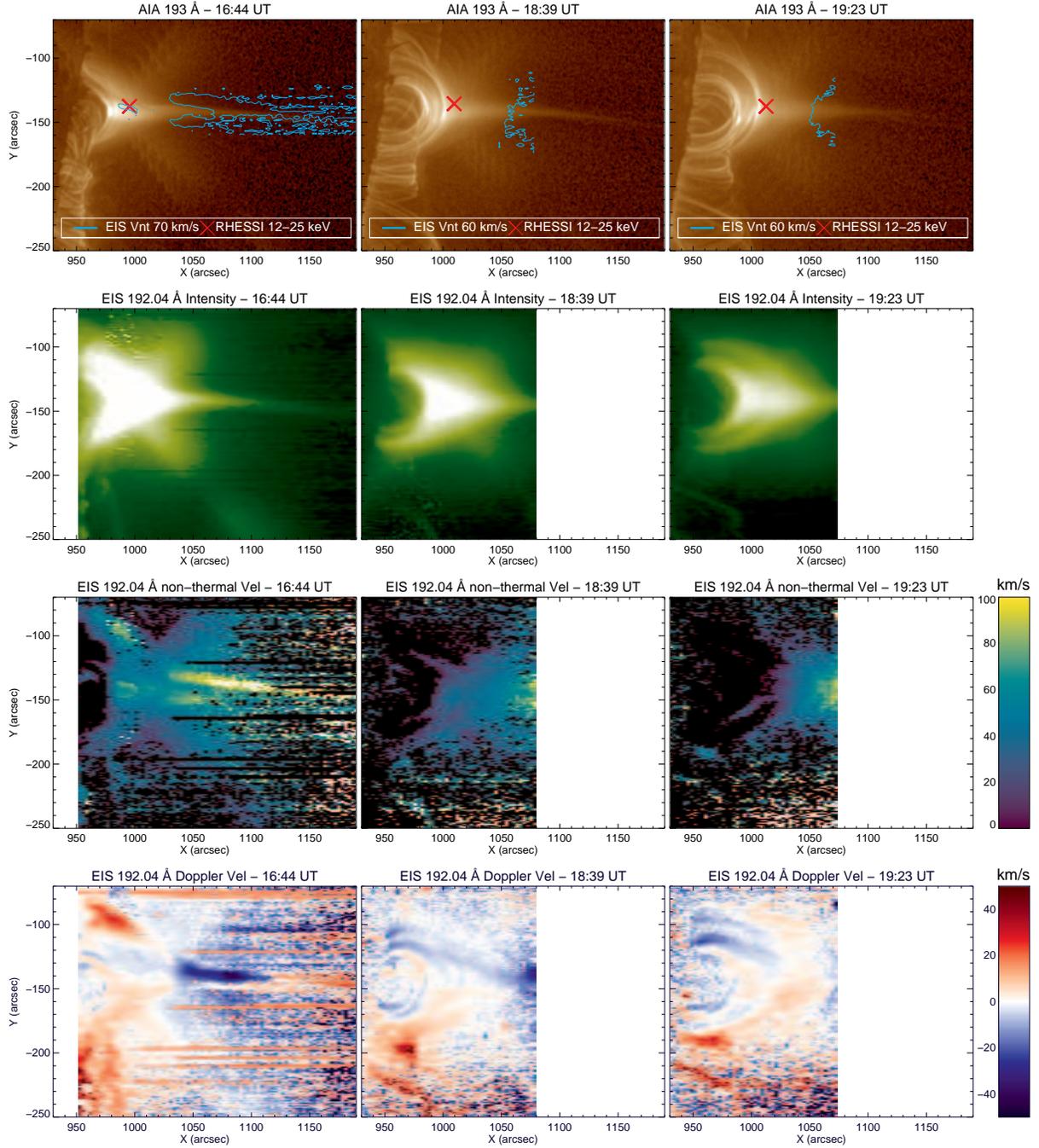}
  \caption{From left to right, the three columns show data at 2017 September 10 16:44, 18:39 and 19:23 UT. 
  Top: MGN sharpened \citep{Morgan} AIA 193 {\AA} observations. Blue contours show regions of highest EIS non-thermal velocities from row 3. The red X marks the location of highest 12--25 keV RHESSI emission. 
  Second row: EIS 192.04 {\AA} intensity measurements. 
  Third row: EIS 192.04 {\AA} non-thermal velocity measurements (km/s). 
  Bottom row: EIS 192.04 {\AA} Doppler velocity measurements (km/s).
  A diffraction pattern is visible in earlier frames, seen as a cross emanating from the brightest region.}
  \label{fig:eis}
\end{figure*}

Hinode EIS observations of the impulsive flare phase are studied in \citet{Warren} and \citet{Cheng}, which look at the early EIS data up to 16:44 UT. Analysis of the later EIS observations, from 18:39 - 19:23 UT, are yet to be published. Although the alignment of the earlier EIS observations are optimal for studying the visible plasma sheet, saturation of the detectors by bright features under the plasma sheet makes spectral analysis of this region, including post-flare loops, impossible. In the later observations however, the flare loops have cooled enough for EIS to measure reliable spectral data during the late flare evolution. Unfortunately, pointing for the latter observations only captures the base of the plasma sheet on the western edge of the frame. 

In Figure \ref{fig:eis}, we examine the evolution of the flare region across three EIS Fe XXIV 192.04 {\AA} rasters, with accompanying broadband AIA 193 {\AA} observations. We present the last EIS raster of the earlier observing sequence (16:44 UT) and the first and last raster of the later observing sequence (18:39 and 19:23 UT). It is worth noting therefore that the three frames are not evenly spaced temporally. The EIS data were prepped using the SolarSoftWare $eis\_prep.pro$ routine, with saturated data points removed using $eis\_sat\_windata$ (this only has an effect in rasters at the start of the flare), and lines fitted using $eis\_auto\_fit$. Wavelength correction due to spacecraft orbital variation is also applied. AIA observations around corresponding EIS times were chosen with an exposure time under 1.5 seconds, to minimize effects of saturation in the flare loops. The AIA 193 {\AA} images are sharpened using the the Multi-Scale Gaussian Normalization (MGN) technique \citep{Morgan}.

The EIS Fe XXIV 192.04 {\AA} line has a formation temperature of 17 MK, whereas the broadband AIA 193 {\AA} channel samples both the higher temperature Fe XXIV and cooler temperature Fe XII lines (with formation temperature at 1.2 MK). At 16:44 UT, there is a clear similarity between the EIS and AIA measurements, with both instruments measuring hot plasma in the bright primary flare loops, the prominent horizontal plasma sheet structure, and dimmer flare loops to the south of the frame. The primary flare loops are still significantly saturated in the EIS observations, causing the visible `X'-shape diffraction pattern. This diffraction pattern is still noticeable at 18:39 UT, but gone by 19:23 UT. A slight diffraction pattern of similar shape is also visible in AIA. 

Later in the flare, differences between EIS and AIA observations of the region become more apparent. In AIA, the plasma sheet is still clearly present at 18:39 UT, with the dimmer cusp-like separatrix structure visible around it. In EIS measurements however, these two structures appear to blend into one, forming the dominant cusp feature. As time passes, the flare loops rise and the cusp structure starts to evolve towards a smoother, perhaps less stressed
structure. The southern flare loops are almost invisible in these later EIS times too, despite still seen in AIA. These loops have cooled enough to be visible in Fe XII lines rather than in Fe XXIV, in contrast to the primary flaring area loops, which still contain significant 17 MK Fe XXIV emitting plasma.
Electron temperatures in this region are explored in section 3.3.

\subsection{Doppler velocity}

Doppler velocities of the plasma sheet's early evolution were studied in \cite{Warren}. Unusual red and blue shifts were reported, but these were attributed to an artifact of the instrument's asymmetric point spread function (PSF). These instrumental artifacts are most common where a sudden sharp change in velocity/intensity is observed. A similar pattern in the plasma sheet is seen in the EIS 16:44 UT Doppler data shown in Figure \ref{fig:eis}. Doppler measurements in the flare loops are not available throughout the early EIS observing sequence, due to high levels of saturation in this region.

In the later EIS observations however (from 18:39 UT onwards), the flare loops are no longer saturated. Additionally, the brighter cusp region means there is no longer sudden intensity/velocity variation across the plasma sheet, meaning the PSF effect is less of a concern. Due to this, we are able to present reliable Doppler velocity measurements of the 2017 September 10 event for the first time.

The nominal wavelength scale was determined from average spectra from a region on the disk. We see in Figure \ref{fig:eis} (at 18:39 UT onwards) that the smallest loops closest to the limb are blue shifted on the southern half of the loop, and red shifted on the northern half, while the higher flare loops and cusp region display opposite behaviour. The blue shifted regions show areas of plasma with a component flowing towards the observer, consistent with plasma flowing down from the top of the newly reconnected flare loops \citep[see cartoon in Figure 1 of][]{Polito}.
This Doppler pattern implies that the blue shifted footpoint is closer to the observer than the red shifted footpoint, giving us an indication of the orientation of the flare loops. The flare loops therefore sit close to the plane of sky (POS), but tilted slightly in either the north or south direction. The low LOS velocities of 10-20 km/s support the fact that the loops lay close to the POS, when compared to downflow velocities of several hundred km/s observed in previous flares \citep{antonucci1984}.

\citet{Fleishman} provide a graphic displaying the orientation of the primary flare loops at reconnection onset. They show the southern footpoint of loops positioned on disk, closer to the observer, determined by the absence of an on-disk hard X-ray signature at the northern footpoint. This configuration is consistent with the observed EIS Doppler velocities in the lower flare loops. However, as further flare loops form (either at different altitudes and/or locations along the LOS), they do so with a different orientation from our LOS, resulting in a northern footpoint closer to the observer. Such an effect could be due to either a slight curve in the east-west polarity inversion line, or to higher altitude loops forming with lower shear.

At the edge of the FOV, at the tip of the cusp and base of the plasma sheet, we observe blue shifted plasma with velocities around 32 km/s in the LOS at 18:39 UT. These Doppler velocities align spatially with the regions of high non-thermal velocities. If the plasma sheet were tilted slightly away from the observer in the east-west direction, then blue shifted plasma at this location is likely a result of the downflow of plasma from the plasma sheet. This tilt is likely small, as LOS velocities are low compared to POS downflows of up to 165 km/s measured in a current sheet by \citet{Savage2010}.

\subsection{Non-thermal broadening}

\begin{figure*}
  \centering
  \includegraphics[width=14cm]{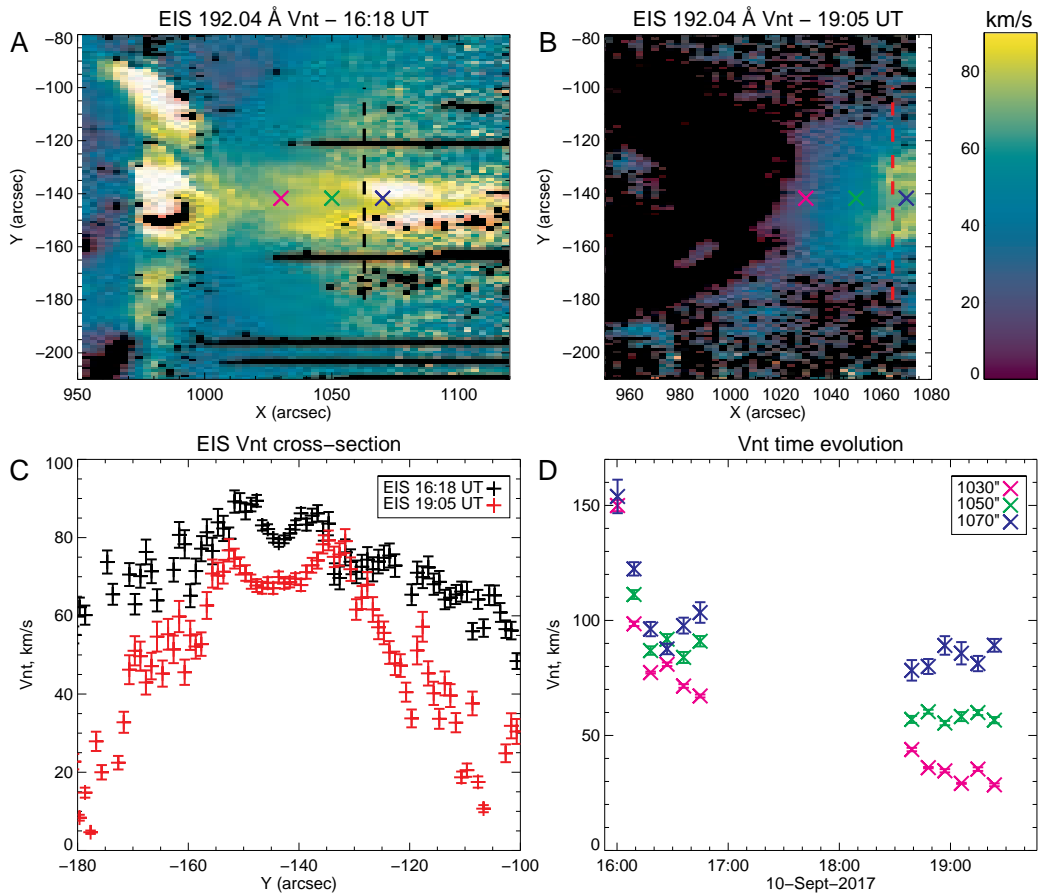}
  \caption{
  A,B: EIS 192.04 {\AA} non-thermal velocity measurements (Vnt) (km/s) for 2017 September 10 16:18 and 19:06 UT respectively. The dashed lines correspond to the cross-sections in panel C, and colored Xs refer to the locations at which we track velocity variations with time in panel D.
  C: Variation in non-thermal velocity across the plasma-sheet at 16:18 and 19:06 UT, along the slice marked in top panels. 
  D: Variation of non-thermal velocity with time, at three locations along the plasma sheet (marked in top panels).}
  \label{fig:vel}
\end{figure*}

Non-thermal broadening is the excess broadening observed in spectral lines after subtracting the instrumental and thermal line-widths. Non-thermal broadening can be caused by several processes, including opacity broadening, Stark broadening, and unresolved plasma flows with no preferential direction in the LOS \citep[e.g.][]{Doschek2014}. However, it is common in the coronal spectroscopy of flares in particular to associate excess line widths with the presence of non-thermal mass motions \citep[e.g. unresolved flows,][]{Antonucci1986} and derive an equivalent RMS speed from the excess broadening. The LOS RMS speed is referred to as the non-thermal velocity, given by $V_{nt} = [2k(T_{D} - T_{e})/m_{i}]^{1/2}$, where $T_{D}$ is the Doppler temperature derived from the total observed line width, $T_{e}$ the electron temperature given by the peak of the contribution function, and $m_{i}$ the mass of the ion considered. Since $T_{e}$ can be verified by line ratios in part of the FOV, and densities during flares are high, the excess line width cannot generally be
accounted for by a difference in ion and electron temperatures.

Recent work by \citet{Kawate2016} does find evidence of departures from a Maxwellian distribution and ionization equilibrium from EIS measurements of highly ionized Fe lines in flares. However, the number of pixels affected in the flares studied were of order 1.4\%, leading them to conclude that the isothermal assumption is valid in most cases given the timescale of EIS exposures. We make this assumption in our own analysis, while acknowledging that departures could exist.

Figure \ref{fig:eis} presents non-thermal velocity maps over the event, with contours of the highest velocities overplotted on the AIA 193 \AA\ images. Both of these show the source of fastest non-thermal velocities rising with time, a continuation of what is seen in the first few EIS rasters of the flare \citep{Warren}. The temporal and spatial evolution of the velocities are shown further in Figure \ref{fig:vel}.

Figure \ref{fig:vel} shows the non-thermal velocity measurements at 16:18 and 19:05 UT. Taking a cross-section of the velocity at a height of $1065 \arcsec$ (Figure \ref{fig:vel}C), we see clearly the bifurcated velocity structure described by \citet{Warren} and \citet{Cheng}. Although the structure is visible in most EIS rasters, it is most prominent in the two shown here. 
This velocity structure is long-lived, with these maps showing data 22 and 189 minutes after the flare onset. In the 166 minute window between these observations, peak velocities only fall from 90 km/s to 80 km/s. As the plot shows however, velocities at the edge of the plasma sheet decrease at a much faster rate. This variation in velocity with time is explored further in Figure \ref{fig:vel}D, where we plot the change in velocity at three different altitudes of 1025, 1050 and $1075 \arcsec$, across the EIS observation times. 

We see non-thermal velocity at these altitudes drop dramatically in the first 20 minutes of the flare, from upwards of 150 km/s down to 100 km/s. From here, velocity falls much slower, with its exact rate dependent upon the altitude. At the lower altitude of $1030 \arcsec$, velocities fall much faster to just 30 km/s at 19:23 UT. At $1050 \arcsec$, velocity remains constant at 60 km/s from 18:39 until the end of EIS observations at 19:23 UT. Velocities are highest around $1070 \arcsec$, still as high as 80 - 90 km/s at the end of EIS observations. They even appear to rise slightly towards the end of the observing sequence.

\subsection{Electron Temperatures}

\begin{figure*}
  \centering
  \includegraphics[width=14cm]{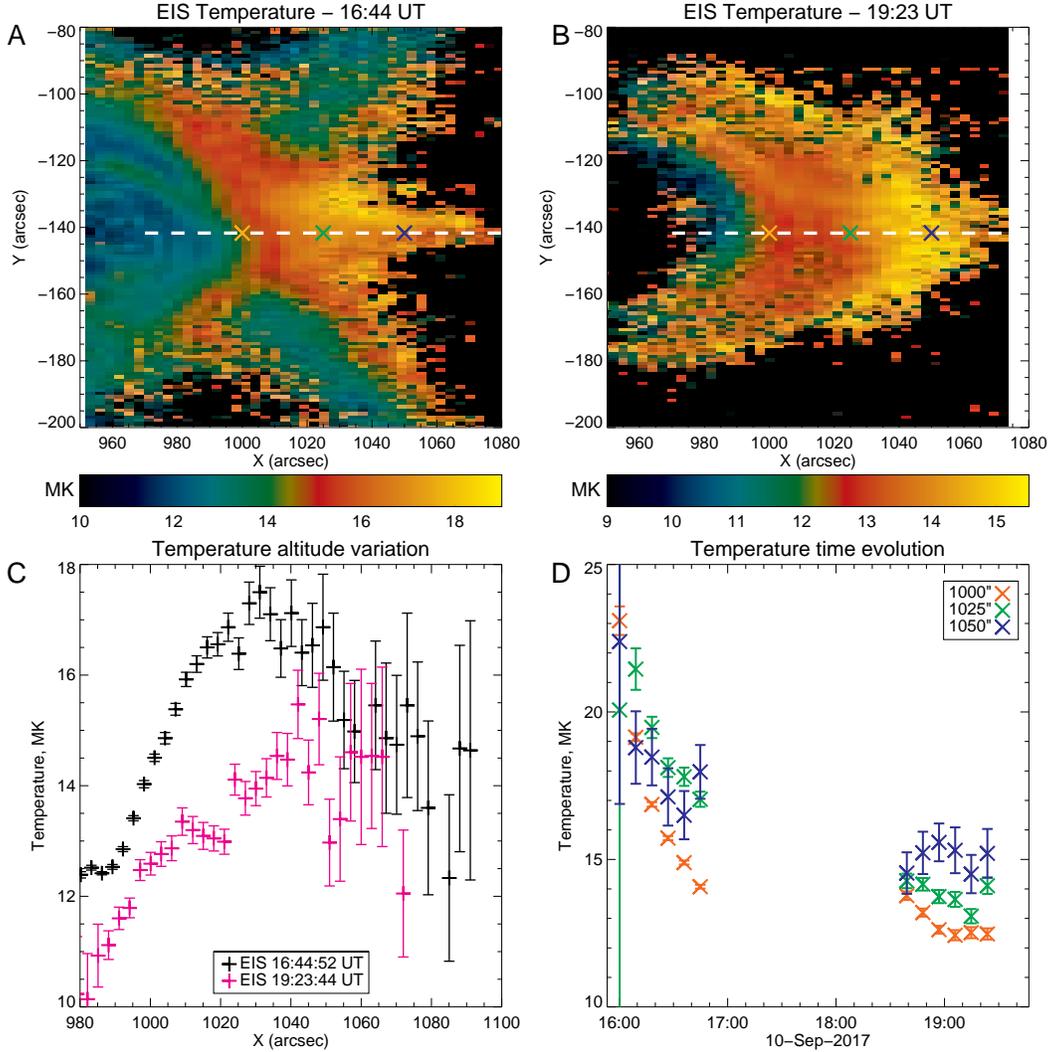}
  \caption{
  A,B: Temperature determined from the ratio of EIS Fe XXIV 255.10 {\AA} to Fe XXIII 263.76 {\AA}, at 2017 September 10 16:44 and 19:23 UT respectively. The dashed white line shows the cross-section slice shown in the panel C. Colored Xs mark the locations of temperature variation plotted in the panel D.
  C: Variation in temperature along the horizontal cross-section at 16:44 and 19:23 UT. Location of cross-section marked above.
  D: Variation of temperature with time, at the three locations marked in top panels.
  }
  \label{fig:Temp}
\end{figure*}

As in \cite{Warren}, we can estimate the electron temperature of the region using the temperature sensitive EIS Fe XXIV 255.10 {\AA} / Fe XXIII 263.76 {\AA} line pair. Emission in these lines does not extend to as high an altitude as Fe XXIV 192.04 {\AA} emission, but is observed near the base of the plasma sheet in the early phase of the flare. Similarly to Fe XXIV 192.04 {\AA}, the central flaring region is saturated, causing a visible diffraction `X' for the duration of the first EIS observing sequence (Figure \ref{fig:Temp}A). The diffraction pattern has disappeared by the later EIS rasters however, and we are able to discern the shape of the flare loops and cusp at the base of the sheet. Qualitatively assessing Figure \ref{fig:Temp}B, we see a rise in temperature across the flare arcade with increasing height. This temperature increase is validated quantitatively in Figure \ref{fig:Temp}C, plotting the temperature profile along the flaring region at 16:44 and 19:23 UT, the last raster of both observing sequences. We measure the temperature up to $1040 \arcsec$ with reasonable uncertainty, but this uncertainty increases further with altitude.

We observe that temperatures across the arcade fall between the two frames, albeit at different rates at different heights. Maximum electron temperature occurs at a projected height of $1030 \arcsec$ at 16:44 UT, but around $1050-1060 \arcsec$ at 19:23 UT, at the top of the cusp region. This observed change of temperature across the arcade profile is consistent with the heating of progressively higher loops along the LOS as the cusp grows throughout the flare's duration. The variation in temperature with time is shown further in Figure \ref{fig:Temp}D, for the three different altitudes marked in the top frames of this figure. We see a rapid drop in temperature during the first thirty minutes of flare, only for the rate of cooling to rapidly decrease. This temperature evolution follows a similar trend to the velocity evolution seen in Figure \ref{fig:vel} and in agreement with work from \citet{Warren}, where the cooling rate of flare loops decreases with time as the densities within them decrease. As with the non-thermal velocity evolution, near constant temperatures can be seen at higher altitudes for the duration of the second EIS observing sequence.

\section{Flare Loops}

\begin{figure*}
  \centering
  \includegraphics[width=16cm]{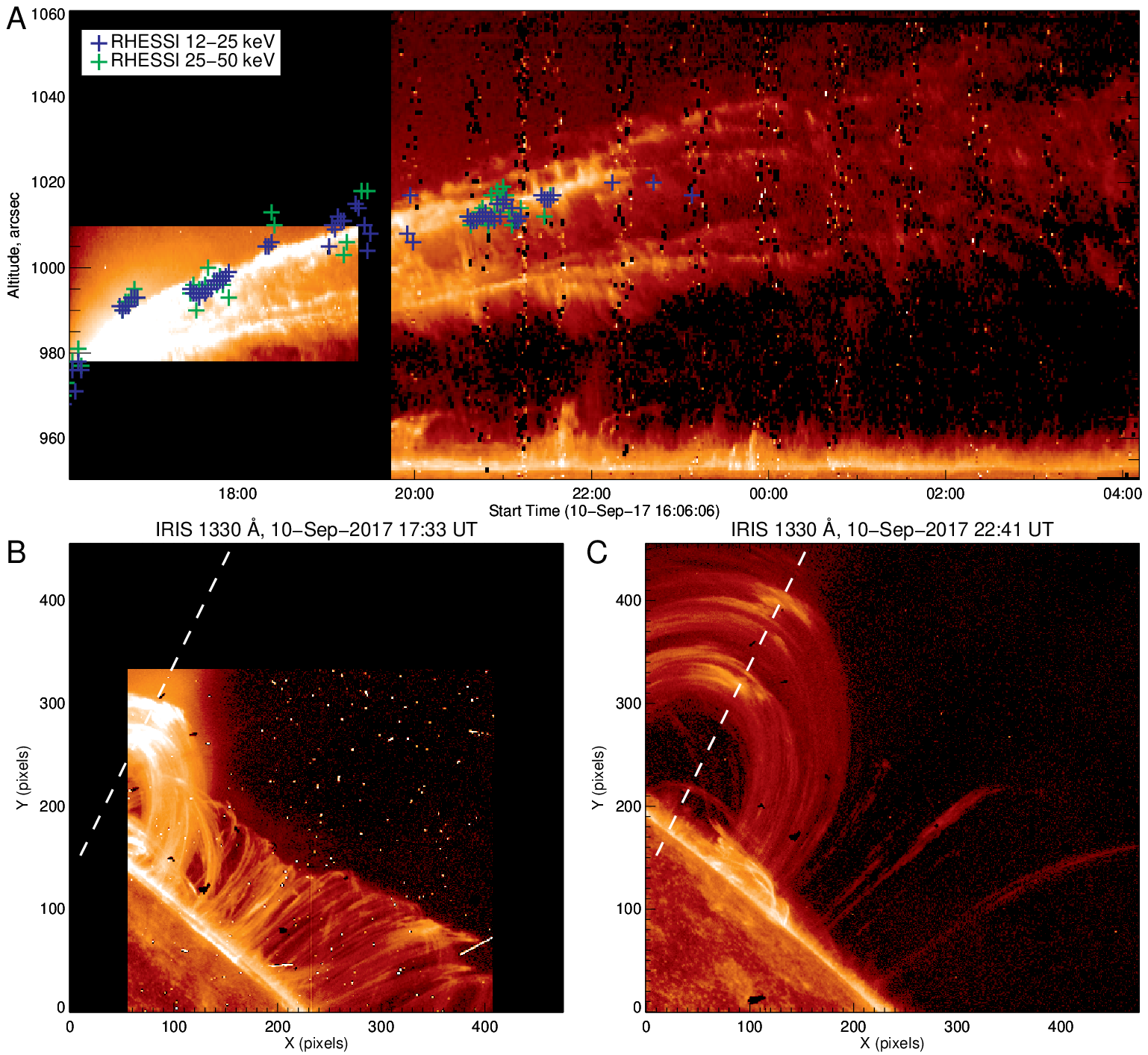}
  \caption{
  A: Height-time plot of a cross-section through the primary flare loop tunnel, as seen in IRIS 1330 {\AA} images. The blue and green crosses mark the corresponding location of RHESSI 12--25 and 25--50 keV emission respectively.
  B,C: Example IRIS 1330 {\AA} \textit{large} and Example IRIS 1330 {\AA} \textit{very large} slit-jaw images with marked cross-section location (white dashed line). Context for the FOV is given in Figure \ref{fig:FOV}.
  }
  \label{fig:IRIS_height}
\end{figure*}

After forming due to magnetic reconnection, individual flare loops shrink rather than expand \citep[e.g.][]{ForbesActon1996}. Therefore, the continued expansion of a flare loop arcade indicates the formation of newly reconnected loops at increasing heights as a result of ongoing reconnection. We can track this growth with IRIS and AIA observations.

At the beginning of the flare, IRIS was centered on the southern arcade and captured the edge of the primary loop tunnel as the CME erupted. A Supra-Arcade Fan (SAF) structure is observed over this southern arcade, the spectra of which are presented in \citet{Cai}.

During the impulsive phase of the eruption, the primary flare loop tunnel (top left of Figure \ref{fig:IRIS_height}B) is seen to grow, as new flare loops form at progressively higher altitudes. The flare loops continue to grow well into the following day (visible in IRIS and AIA observations), despite the active region rotating further around the limb.

Figure \ref{fig:IRIS_height} B and C show a cross-section through the primary flare loops, in both the earlier \textit{large} (12:00 - 19:22 UT) and \textit{very large} (19:44 UT onwards) IRIS 1330 {\AA} slitjaw observing sequences, showing flare loops containing plasma already cooled to around $25000$ K. The orientation of these frames are shown in Figure \ref{fig:FOV}. By plotting intensity along this cross section, we produce the height--time plot shown in Figure \ref{fig:IRIS_height}A. From this diagram, we can see the growth of the brightest flare loops, with the formation of additional dimmer loops above. The bulk expansion of the flare loop system is fast at first, and slows to an estimated 1 km/s growth rate after $\approx$ 20:00 UT. In IRIS observations, the height of the flare loops peak around $90 \arcsec$ above the limb at 01:00 UT on the morning of 2017 September 11, around 9 hours after the flare onset and CME eruption. 

The growth is even more remarkable considering the rotation of the active region onto the far side of the Sun, eclipsing the base of the flare loops. Using simple trigonometry with the solar rotation rate, we estimate that the rotation of the arcade over the western limb decreases the expansion speed by 10\% during the time period analyzed. The bulk growth of a similar flare-loop arcade was analyzed by \citet{vanDriel1997}, who also found an arcade growth rate of 1 km/s in the late phase of an X-class flare.

\subsection{RHESSI Looptop sources}

According to the standard flare model, HXR emission can be produced at the flare looptops as high energy particles outflow from the reconnecting current sheet and collide with cooler plasma in the newly formed flare loops. The cartoon in Figure 2 of \citet{Chen15} demonstrates this process, mirroring the observations in the top row of Figure \ref{fig:eis} in this paper. In Figure \ref{fig:eis}, the red cross shows the location of 12--25 KeV HXR emission as measured by RHESSI \citep{Lin} at the time of each EIS raster, located at the top of the flare loops in each case. Above this, we see the contour of high non-thermal velocity, perhaps a signature of the reconnection outflows shown at the base of the current sheet in the cartoon of \citet{Chen15}.

The early evolution of RHESSI HXR sources with time is presented by \citet{Hayes}. They examine the rise in 6--12 and 12--25 keV HXR signatures from the flare onset at 15:49 to 18:30 UT. In Figure \ref{fig:IRIS_height}, we overplot the location of 12--25 keV emission onto the IRIS height-time plot, from flare onset up to 23:00 UT. We introduce the 25--50 keV source location too, from onset to 21:30 UT. By binning the data over 120 seconds (30 spacecraft rotations), we are able to increase our signal to noise (S/N) ratio and resolve the emission source this late in the flare using the CLEAN algorithm. Binning in this way provides fewer data points, but shows a clear source of 12--25 and 25--50 keV emission 7 and 5.5 hours after the flare onset respectively. 

In Figure \ref{fig:IRIS_height}, we see the HXR source height rising with time, at progressively slower speeds. At later times (around 20:00 UT) the X-ray point source continues to rise with the brightest flare loop, despite newer, fainter loops forming above it.

\begin{figure*}
  \centering
  \includegraphics[width=16cm]{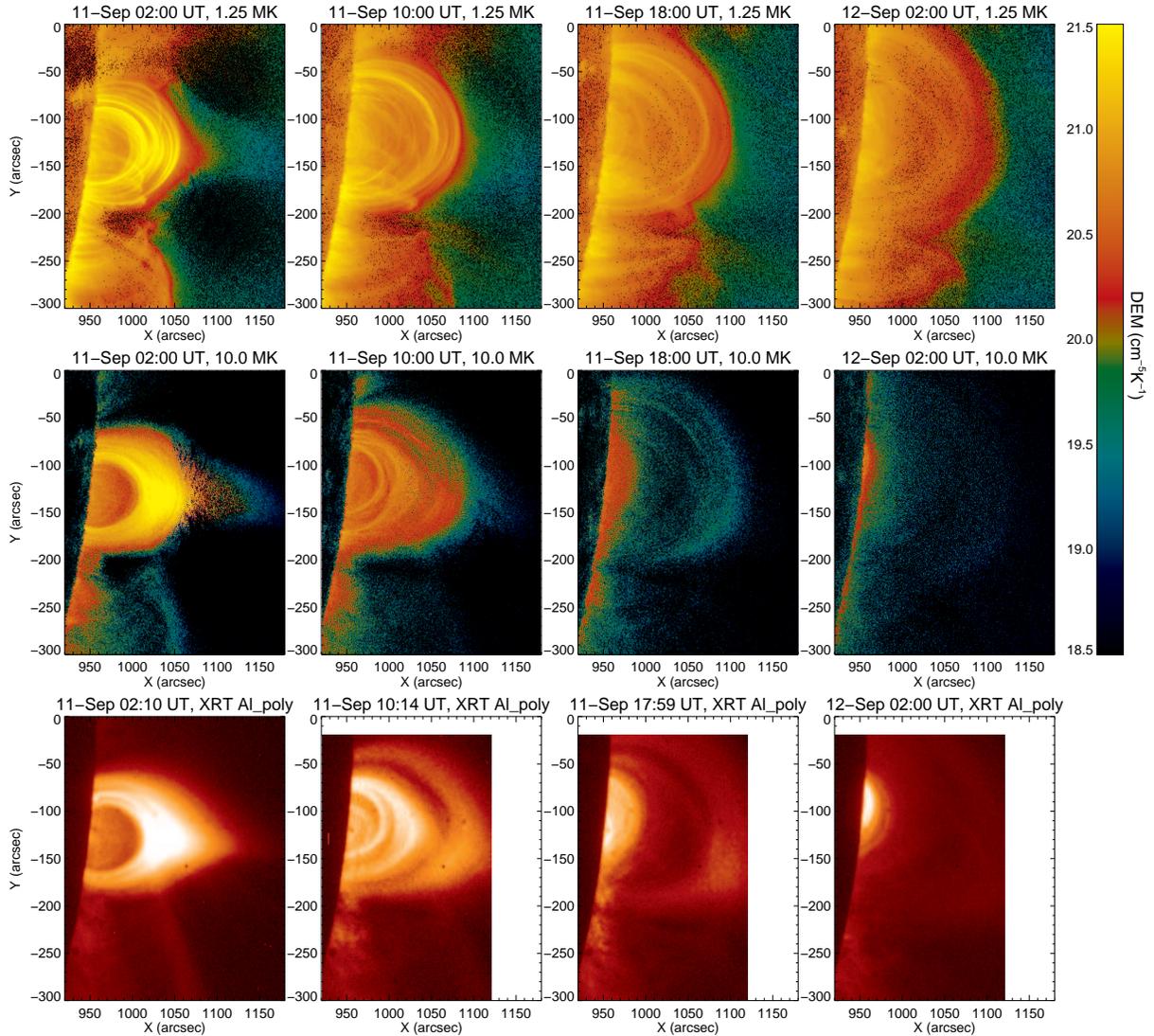}
  \caption{
  Top: Time evolution of AIA 1.25 MK DEMs. 
  Middle: Time evolution of AIA 10 MK DEMs. 
  Bottom: Time evolution of XRT Al poly.
  }
  \label{fig:DEMs}
\end{figure*}

\subsection{Continued loop growth}

After the off-limb IRIS 1330 {\AA} signal becomes too weak to track loop growth past September 11 01:00 UT, we see the loops continue to rise in AIA images. Differential Emission Measures (DEMs) can be used to track this growth at different temperatures, and Figure \ref{fig:DEMs} shows the evolution of Emission Measure (EM) for 1.25 and 10.0 MK plasma, \citep[constructed using the DEM code of][]{Hannah}. The top row of Figure \ref{fig:DEMs} shows the growth of the loops in cooler plasma, where both the primary (east-west) and southern (north-south) arcades continue to grow in height until around 12 Sep 02:00 UT -- 34 hours after the flare onset. This continued growth is despite the fact that the active region is rotating further over the western limb, meaning a structure of constant size would appear to shrink due to projection effects. 

We see a similar effect with the hotter plasma too. On September 11 at 02:00 UT, 10 hours after the flare onset, the base of the plasma sheet is still visible at these high temperatures. The flare loops also have a more `cusp-like' shape, in comparison to the relaxed lower-temperature loops \citep{Gou2015}. This cusp structure is also visible in XRT, where the Al poly observations are most sensitive to plasma at around 6.9 MK. This cusp feature seems to tilt to the south over the next 8 hours, where we see a small plasma sheet signal visible in the high temperature DEMs at 11 Sep 10:00 UT. The XRT FOV has shifted at this time, but the loop structure in these images still match those in the high temperature DEMs; visibly different to the loops seen at the same location in cooler plasma. To investigate this further, future study could investigate the flare loop cooling rate using DEMs, comparing with theoretical cooling rates. Although the EM of high temperature plasma grows weaker, there is still some plasma emitting at high temperatures on 12 Sep at 02:00 UT, as the loops reach their maximum observed height as they are eclipsed by the limb. 

\section{Polarization Measurements}

\begin{figure*}
  \centering
  \includegraphics[width=16cm]{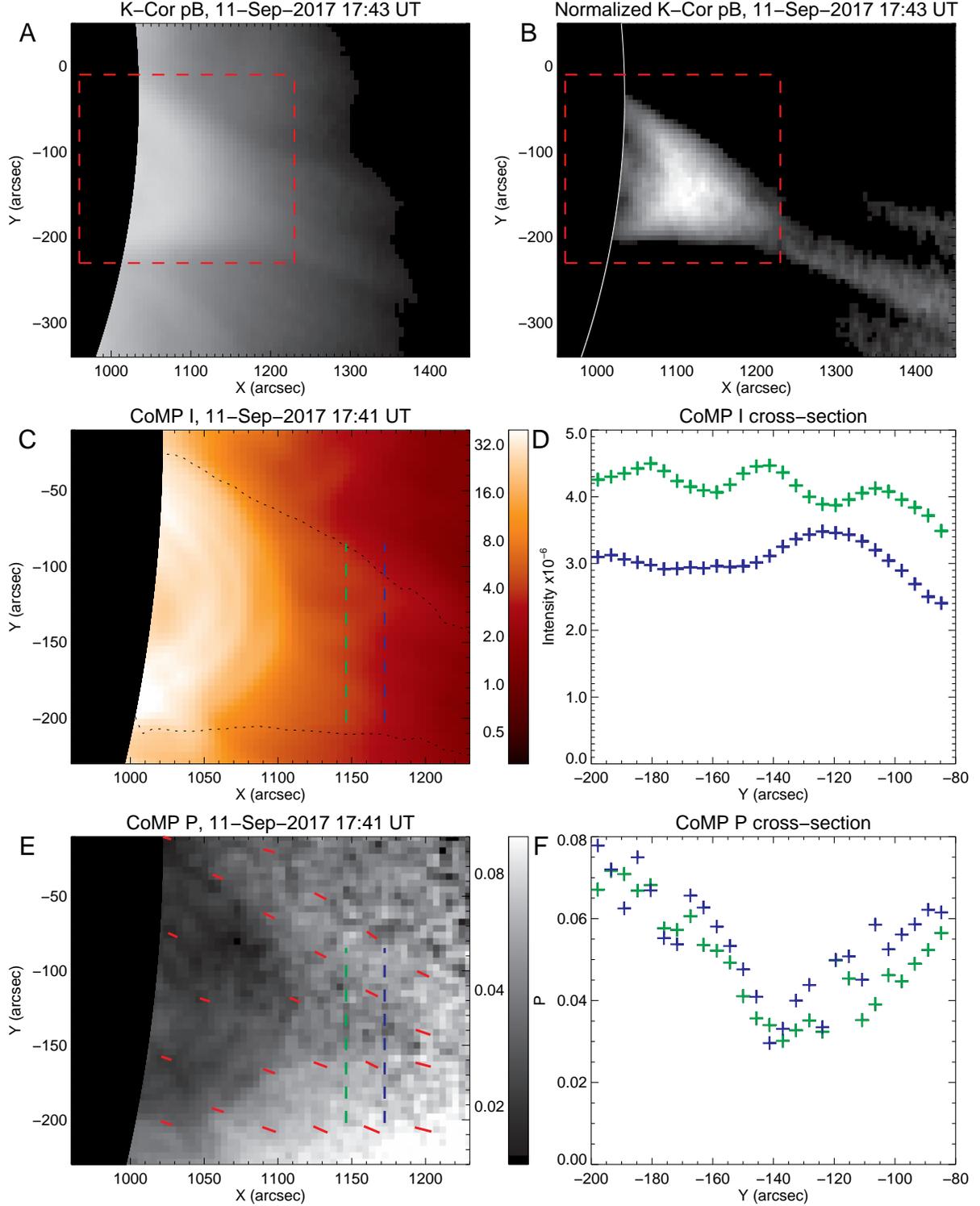}
  \caption{
  A: K-Cor polarized brightness, pB. Red dashed box shows FOV of CoMP panels.
  B: K-Cor pB, normalized by mean brightness at each altitude.
  C: CoMP 1074.7 intensity. The black dotted contour marks the location of K-Cor cusp feature in panel B.
  D: Cross-section of CoMP I, at locations marked in panel C with corresponding colors.
  E: CoMP normalized linear polarization, $P=\sqrt{Q^2+U^2}/I$. Red lines are polarization vectors, with length proportional to $-1/\log(P)$.
  F: Cross-section of CoMP P, at locations marked in panel E with corresponding colors.
  }
  \label{fig:CoMP}
\end{figure*}

According to the standard model, the presence of a cusp would imply the existence of a reconnecting current sheet above it. As seen in Figure \ref{fig:DEMs}, there is no evidence of this in EUV images beyond around September 11 10:00 UT. Since the EUV lines are collisionally excited, the EUV intensity falls off more rapidly with height than the infrared forbidden lines observed by CoMP which include a significant radiative contribution. Similarly, the Thomson scattered emission observed by K-Cor falls off less rapidly with height. To detect the remnants of the diffuse plasma sheet above our cusp, we turn to polarized infrared (CoMP) and white light (K-Cor) data.

Figure \ref{fig:CoMP}A shows white light polarized brightness (pB) measurements from K-Cor, time averaged over an hour from 2017 September 11 17:43 UT. K-Cor observes photospheric photons scattered by electrons in the corona and measures a variation of brightness with electron density $n_e$ (rather than $n_e^2$ for EUV images) as, unlike EUV, it does not rely on collisional excitation \citep{Landi2016}. Due to this difference in density dependency, coronal brightness drops more slowly with height in white-light and infrared observations, allowing K-Cor to observe more diffuse coronal structures at much higher altitudes. Studying \ref{fig:CoMP}A, we can see the outline of the bright cusp-like feature on the limb, but it is difficult to resolve any detail. Figure \ref{fig:CoMP}B shows a processed version of this image, normalized by dividing each pixel brightness by the mean brightness at the corresponding altitude within the FOV. The plasma sheet is clearly visible and emanating from the cusp top towards the south-west. This direction was also indicated from the apparent orientation of the cusp consistent with XRT data shown in Figure \ref{fig:DEMs}. This southward deflection is interesting, given that the heliospheric current sheet is located in the northern hemisphere at this solar longitude. This deflection is then perhaps due to the over-expansion of open field above a coronal hole to the north of the flare loops (but to the south of the heliospheric current sheet). Projection effects may also contribute to this southward deflection as the flaring region rotates behind the limb. Although the plasma sheet is no longer visible in LASCO at this time, the instrument has observed a similar deflection of CMEs in the past \citep{Kay2017}.

\subsection{Coronal Multi-channel Polarimeter}

Using infrared polarization measurements from CoMP, we can examine more closely the base of the plasma sheet structure observed by K-Cor. Linear polarization provides unique and complementary 
information about source regions, enabling the study of thermal and magnetic conditions of both large-scale \citep[][]{Dove2011,Gibson} and small-scale \citep[][]{French} structures in the solar corona. For the Fe XIII emission observed by CoMP, linear polarization (normalized by measured intensity, $P = \sqrt{U^2 + Q^2}/I$) is proportional to $P \propto 3 \cos^2\theta_B-1$ \citep{Casini,Judge2007,French}, where $\theta_B$ is the POS angle between the center of the incident solar radiation and local magnetic field. Therefore, $P$ varies between a maximum value at $\theta_B=0$ (i.e. a radial field), to a minimum of zero at $\theta_B=\theta_{VV} = 54.74^{\circ}$, where $\theta_{VV}$ is the `Van Vleck' angle. 
The maximum $P$ value at $\theta_B=0$ is determined by the density and temperature of the emitting plasma, and the local anisotropy of incident radiation from the solar disk.
$\theta_B$ can be estimated with a 90$^{\circ}$ ambiguity through $\theta = \frac{1}{2}\arctan{(U/Q)}$, where $\theta$ is the polarization angle.

In \citet{French}, CoMP observations of the September 10 flare revealed a region of low polarization ($P\approx 0.01$), aligned with the plasma sheet in EUV. Such a small $P$ value cannot result from depolarizing particle collisions, as density effects are not sufficient. \citet{French} concluded that depolarization occurs through unresolvably small variations in $\theta_B$. Comparing the data with exploratory models using the Coronal Line Emission \citep[CLE;][]{Judge+Casini2001} code, \citet{French} found the polarization signature is most likely a result of reconnection turbulence or plasmoid instabilities, four hours after the flare onset.

Figure \ref{fig:CoMP} presents CoMP data from 2017 September 11, averaged over 8 images between 17:41 -- 18:47 UT. The Fe XIII 1074.7 nm line observed by CoMP has a formation temperature of 1-2 MK \citep{Judge2010}, mapping well to the 1.25 MK EM plasma shown in Figure \ref{fig:DEMs}. The CoMP intensity ($I$) maps shows the flare loops above the occulting disk, which were only just visible 24 hours earlier in the September 10 observations \citep{French}. The K-Cor cusp shape is overlaid on CoMP $I$ in Figure \ref{fig:CoMP}C. Unlike white light $pB$, CoMP $I$ only observes the 
brighter flare loops
with electron temperatures favorable to Fe XIII emission. Taking a cross-section in two locations above the flare loops (Figure \ref{fig:CoMP}D), we do not see any clear structure. This observation is compatible with the lack of a plasma sheet signature in CoMP $I$ on 2017 September 10, despite its presence in EUV and CoMP $P$.

CoMP $P$ is shown in Figure \ref{fig:CoMP}E. Within the bright post-flare loops we have a low linear polarization structure, likely a result of depolarization due to collisions with thermal electrons and protons. Although there is slight variation of $P$ along the loops, the lack of dependency on field direction (assuming the loops trace the magnetic field), indicates a mix of $\theta_B$ along the LOS down the flare loop arcade.
Above these obvious post-flare loops, we see a subtle, darker triangle-like structure, aligned to the center of the bright cusp feature in K-Cor $pB$ data. The structure is also consistent with the last EUV observations made of the plasma sheet, in 10 MK DEMs (Figure \ref{fig:DEMs}) on September 11 at 02:00 and 10:00 UT. Taking a cross-section through this darker region, we find a minimum $P$ value of 0.03, from a maximum of 0.06 and 0.075 to the north and south of the feature respectively. Two sample cross-sections are shown in Figure \ref{fig:CoMP}F. As seen in the K-Cor data, the polarization vectors show that the field in the region is deflected towards the south. This deflection explains the asymmetry in the $P$ cross-section, as southwards-pointing fields are orientated further from the radial direction at the north of the plasma sheet, resulting in a lower polarization value. 

Using the density sensitive intensity ratio of Fe XIII 1079.8 to 1074.7 nm \citep{French} and taking into account
the contributions from 
radiative excitation with CLE, 
we estimate the density of 1-2 MK plasma above the flare loops (at $1140 \arcsec$) to be $1.3\times 10^8$ cm$^{-3}$. In the associated CLE calculations, we built a new plasma sheet model that 
includes 
a laminar field aligned parallel and anti-parallel to the the plasma sheet observed by K-Cor, as in \citet{French}. Using the density estimate at $1140 \arcsec$, the model yields a minimum polarization of $P \approx 0.05$ in the plasma sheet. This is greater than the minimum of $P \approx 0.03$ observed in the September 11 CoMP observations. Just as on September 11, we interpret this low $P$ value to be a result of structured magnetic field (varying $\theta_B$) in the LOS and POS on sub-pixel scales \citep{French}. The increase of polarization 
between 10 and 11 September is consistent with continued presence of magnetic reconnection instabilities, embedded in
a more laminar magnetic configuration as the system loses free magnetic energy.

\section{Discussion}
We have analysed previously unpublished EIS data from three hours after the flare onset at 15:44 UT. Similarly to EIS observations from the impulsive phase, we find high non-thermal velocities at the base of the plasma sheet, consistent with the presence of sub-resolution 
plasmoid-modulated reconnection or turbulence \citep{Warren}. \citep[Although alternative interpretations of the non-thermal velocities are possible, this interpretation is consistent with additional evidence presented in][] {Cheng,French}.
These velocities, 80 km/s at 19:23 UT, are only marginally lower than the velocities early in the flare at 16:18 UT, implying that the non-thermal processes are active at a similar level at this time. As noted in \citet{Polito2015} and \citet{Polito}, enhanced line widths can also be produced by a rapid increase in ion temperatures. Interpreting the excess widths measured by EIS in these terms would imply an effective ion temperature in the range of 40 MK, which is more than twice the Fe XXIV peak formation temperature and electron temperatures measured using line ratios in the lower plasma sheet at this time. This assumes that ion and electron temperatures are equal in this case, which, as discussed earlier may not always be true \citep{Kawate2016}. Work by \citet{Dudik2016} and \citet{Polito2018} demonstrates that non-equilibrium ionization and non-thermal particle distributions may lead to ions being formed at much higher temperatures than expected, which could in turn contribute to a broader line profile. Further analysis of the line profiles and their variation with position and magnetic field orientation may help in quantifying the contribution of such effects.

These non-thermal velocities, combined with the low polarization structure in 2017 September 10 CoMP measurements presented in \citet{French}, suggests that significant reconnection instabilities are likely still present on 2017 September 10 at 20:00 UT. Additional observations during this period are also consistent with signatures of impulsive energy release, within the context of the standard eruptive flare model. As we look down the flare loop tunnel under the erupting flux rope \citep{Chen20}, we see RHESSI HXR signatures at the loop tops as high energy particles from the plasma sheet above collide with the closed loops. We also find Doppler signatures in EIS measurements of down-flowing (presumably cooling) plasma along the reconnected loops, with subsequent loops forming at increasing altitudes as reconnection continues to take place.

Using RHESSI HXR, high resolution IRIS imaging, AIA and XRT, we are able to track the growth of flare loops with time
from plasma with 
diverse thermal conditions,
establishing that the flare evolution displays the classical signatures of cooler plasma in more relaxed loops, and hotter plasma in a tighter cusp shape. Although signatures of ongoing reconnection many hours after flare onset have been seen before \citep[e.g.][]{Savage2010}, the observations here indicate that this cusp feature is still seen with a plasma sheet emanating from its looptop, up to 16 hours after the flare onset, with significant plasma still at temperatures of 10 MK. The presence of these structures, together with the persistent presence of HXR emission at high altitudes is a strong implication that fast magnetic reconnection is still occurring in this case. Observations of long-lived $\gamma$-ray emission observed by FERMI provide evidence of particle acceleration over similar timescales \citep[12 hours,][]{Omodei}, consistent with this interpretation. 

On even longer timescales, we examined K-Cor pB and CoMP linear polarization measurements from over a day into the flare's evolution. At this time, K-Cor observed a much larger cusp region and apparent plasma sheet deflected towards the south (the direction of which is also seen in earlier DEM measurements). The EUV and white-light plasma sheet structure is also consistent with linear polarization measurements from CoMP at 2017 September 11 17:41 UT, where low levels of polarization at the base of the suspected sheet can only be accounted for by sub-pixel magnetic structure in the LOS and POS, implying the continued presence of magnetic reconnection. The fact that infrared linear polarization measurements can detect signatures of reconnection in this way, despite no visual signatures in EUV, has exciting implications for off-limb flare observations from the DKIST CRYO-NIRSP instrument. It emphasizes the need to complement EUV capabilities with optical/IR instruments capable of measuring structure from the solar surface to a few solar radii.

\section{Conclusion}
The 2017 September 10 flare has been extremely well studied, but most studies have so far focused primarily on the impulsive phase of the event. In this work we have shown that the dynamical energy release signatures and flare configuration associated with the impulsive phase of the standard eruptive flare model are still present many hours into the flare's evolution, indicating that in this case fast reconnection is likely still occurring at least 27 hours after the flare onset. This interpretation supports the results of \citet{French}, which indicated that the presence of magnetic sub-structure within the main plasma sheet is consistent with the presence of unresolvable plasmoids and/or turbulence, only 4 hours after the flare onset. These observations suggest that there is still much to be learned from the evolution of what is quickly becoming one of the most-studied solar flares of all time.

\acknowledgments
R.J.F. thanks the STFC for support via funding from the PhD Studentship, as well as NCAR for funding visits to the High Altitude Observatory via the Newkirk
Fellowship. 
S.A.M and L.v.D.G. are partially funded under STFC consolidated grant number ST/N000722/1. L.v.D.G. acknowledges the Hungarian National Research, Development and Innovation Office grant OTKA K-113117.
DML is grateful to the Science Technology and Facilities Council for the award of an Ernest Rutherford Fellowship (ST/R003246/1). CoMP and K-Cor data services were provided by MLSO. Hinode is a Japanese mission developed and launched by ISAS/JAXA, collaborating with NAOJ as a domestic partner, NASA and STFC (UK) as international partners. Scientific operation of the Hinode mission is conducted by the Hinode science team organized at ISAS/JAXA. This team mainly consists of scientists from institutes in the partner countries. Support for the post-launch operation is provided by JAXA and NAOJ (Japan), UKSA (U.K.), NASA (U.S.A.), ESA, and NSC (Norway).
This material is based upon work supported by the National Center for Atmospheric Research, which is a major facility sponsored by the National Science Foundation under Cooperative Agreement No. 1852977

\bibliographystyle{aasjournal}
\bibliography{bibliography}

\begin{thebibliography}{}
\expandafter\ifx\csname natexlab\endcsname\relax\def\natexlab#1{#1}\fi
\providecommand{\url}[1]{\href{#1}{#1}}

\bibitem[{{Antonucci} {et~al.}(1984){Antonucci}, {Gabriel}, \&
  {Dennis}}]{antonucci1984}
{Antonucci}, E., {Gabriel}, A.~H., \& {Dennis}, B.~R. 1984, \apj, 287, 917

\bibitem[{{Antonucci} {et~al.}(1986){Antonucci}, {Rosner}, \&
  {Tsinganos}}]{Antonucci1986}
{Antonucci}, E., {Rosner}, R., \& {Tsinganos}, K. 1986, \apj, 301, 975

\bibitem[{{Aulanier} {et~al.}(2013){Aulanier}, {D{\'e}moulin}, {Schrijver},
  {Janvier}, {Pariat}, \& {Schmieder}}]{Aulanier2013}
{Aulanier}, G., {D{\'e}moulin}, P., {Schrijver}, C.~J., {et~al.} 2013, \aap,
  549, A66

\bibitem[{{Aulanier} {et~al.}(2012){Aulanier}, {Janvier}, \&
  {Schmieder}}]{Aulanier2012}
{Aulanier}, G., {Janvier}, M., \& {Schmieder}, B. 2012, \aap, 543, A110

\bibitem[{{Bruzek}(1964)}]{Bruzek64}
{Bruzek}, A. 1964, \apj, 140, 746

\bibitem[{{Cai} {et~al.}(2019){Cai}, {Shen}, {Raymond}, {Mei}, {Warmuth},
  {Roussev}, \& {Lin}}]{Cai}
{Cai}, Q., {Shen}, C., {Raymond}, J.~C., {et~al.} 2019, \mnras, 489, 3183

\bibitem[{{Carmichael}(1964)}]{Carmichael}
{Carmichael}, H. 1964, NASA Special Publication, 50, 451

\bibitem[{{Casini} \& {Judge}(1999)}]{Casini}
{Casini}, R., \& {Judge}, P.~G. 1999, \apj, 522, 524

\bibitem[{{Chen} {et~al.}(2015){Chen}, {Bastian}, {Shen}, {Gary}, {Krucker}, \&
  {Glesener}}]{Chen15}
{Chen}, B., {Bastian}, T.~S., {Shen}, C., {et~al.} 2015, Science, 350, 1238

\bibitem[{{Chen} {et~al.}(2020){Chen}, {Yu}, {Reeves}, \& {Gary}}]{Chen20}
{Chen}, B., {Yu}, S., {Reeves}, K.~K., \& {Gary}, D.~E. 2020, \apjl, 895, L50

\bibitem[{{Cheng} {et~al.}(2018){Cheng}, {Li}, {Wan}, {Ding}, {Chen}, {Zhang},
  \& {Liu}}]{Cheng}
{Cheng}, X., {Li}, Y., {Wan}, L.~F., {et~al.} 2018, \apj, 866, 64

\bibitem[{{Culhane} {et~al.}(2007){Culhane}, {Harra}, {James}, {Al-Janabi},
  {Bradley}, {Chaudry}, {Rees}, {Tandy}, {Thomas}, {Whillock}, {Winter},
  {Doschek}, {Korendyke}, {Brown}, {Myers}, {Mariska}, {Seely}, {Lang}, {Kent},
  {Shaughnessy}, {Young}, {Simnett}, {Castelli}, {Mahmoud}, {Mapson-Menard},
  {Probyn}, {Thomas}, {Davila}, {Dere}, {Windt}, {Shea}, {Hagood}, {Moye},
  {Hara}, {Watanabe}, {Matsuzaki}, {Kosugi}, {Hansteen}, \&
  {Wikstol}}]{Culhane}
{Culhane}, J.~L., {Harra}, L.~K., {James}, A.~M., {et~al.} 2007, \solphys, 243,
  19

\bibitem[{{De Pontieu} {et~al.}(2014){De Pontieu}, {Title}, {Lemen}, {Kushner},
  {Akin}, {Allard}, {Berger}, {Boerner}, {Cheung}, {Chou}, {Drake}, {Duncan},
  {Freeland}, {Heyman}, {Hoffman}, {Hurlburt}, {Lindgren}, {Mathur}, {Rehse},
  {Sabolish}, {Seguin}, {Schrijver}, {Tarbell}, {W{\"u}lser}, {Wolfson},
  {Yanari}, {Mudge}, {Nguyen-Phuc}, {Timmons}, {van Bezooijen}, {Weingrod},
  {Brookner}, {Butcher}, {Dougherty}, {Eder}, {Knagenhjelm}, {Larsen},
  {Mansir}, {Phan}, {Boyle}, {Cheimets}, {DeLuca}, {Golub}, {Gates}, {Hertz},
  {McKillop}, {Park}, {Perry}, {Podgorski}, {Reeves}, {Saar}, {Testa}, {Tian},
  {Weber}, {Dunn}, {Eccles}, {Jaeggli}, {Kankelborg}, {Mashburn}, {Pust},
  {Springer}, {Carvalho}, {Kleint}, {Marmie}, {Mazmanian}, {Pereira}, {Sawyer},
  {Strong}, {Worden}, {Carlsson}, {Hansteen}, {Leenaarts}, {Wiesmann},
  {Aloise}, {Chu}, {Bush}, {Scherrer}, {Brekke}, {Martinez-Sykora}, {Lites},
  {McIntosh}, {Uitenbroek}, {Okamoto}, {Gummin}, {Auker}, {Jerram}, {Pool}, \&
  {Waltham}}]{DePontieu}
{De Pontieu}, B., {Title}, A.~M., {Lemen}, J.~R., {et~al.} 2014, \solphys, 289,
  2733

\bibitem[{{Doschek} {et~al.}(2014){Doschek}, {McKenzie}, \&
  {Warren}}]{Doschek2014}
{Doschek}, G.~A., {McKenzie}, D.~E., \& {Warren}, H.~P. 2014, \apj, 788, 26

\bibitem[{{Dove} {et~al.}(2011){Dove}, {Gibson}, {Rachmeler}, {Tomczyk}, \&
  {Judge}}]{Dove2011}
{Dove}, J.~B., {Gibson}, S.~E., {Rachmeler}, L.~A., {Tomczyk}, S., \& {Judge},
  P. 2011, \apjl, 731, L1

\bibitem[{{Dud{\'\i}k} {et~al.}(2016){Dud{\'\i}k}, {Polito}, {Janvier},
  {Mulay}, {Karlick{\'y}}, {Aulanier}, {Del Zanna}, {Dzif{\v{c}}{\'a}kov{\'a}},
  {Mason}, \& {Schmieder}}]{Dudik2016}
{Dud{\'\i}k}, J., {Polito}, V., {Janvier}, M., {et~al.} 2016, \apj, 823, 41

\bibitem[{Fleishman {et~al.}(2020)Fleishman, Gary, Chen, Kuroda, Yu, \&
  Nita}]{Fleishman}
Fleishman, G.~D., Gary, D.~E., Chen, B., {et~al.} 2020, Science, 367, 278

\bibitem[{{Forbes} \& {Acton}(1996)}]{ForbesActon1996}
{Forbes}, T.~G., \& {Acton}, L.~W. 1996, \apj, 459, 330

\bibitem[{{French} {et~al.}(2019){French}, {Judge}, {Matthews}, \& {van
  Driel-Gesztelyi}}]{French}
{French}, R.~J., {Judge}, P.~G., {Matthews}, S.~A., \& {van Driel-Gesztelyi},
  L. 2019, \apjl, 887, L34

\bibitem[{{Gibson} {et~al.}(2017){Gibson}, {Dalmasse}, {Rachmeler}, {De Rosa},
  {Tomczyk}, {de Toma}, {Burkepile}, \& {Galloy}}]{Gibson}
{Gibson}, S.~E., {Dalmasse}, K., {Rachmeler}, L.~A., {et~al.} 2017, \apjl, 840,
  L13

\bibitem[{{Golub} {et~al.}(2007){Golub}, {Deluca}, {Austin}, {Bookbinder},
  {Caldwell}, {Cheimets}, {Cirtain}, {Cosmo}, {Reid}, {Sette}, {Weber},
  {Sakao}, {Kano}, {Shibasaki}, {Hara}, {Tsuneta}, {Kumagai}, {Tamura},
  {Shimojo}, {McCracken}, {Carpenter}, {Haight}, {Siler}, {Wright}, {Tucker},
  {Rutledge}, {Barbera}, {Peres}, \& {Varisco}}]{Golub}
{Golub}, L., {Deluca}, E., {Austin}, G., {et~al.} 2007, \solphys, 243, 63

\bibitem[{{Gou} {et~al.}(2015){Gou}, {Liu}, \& {Wang}}]{Gou2015}
{Gou}, T., {Liu}, R., \& {Wang}, Y. 2015, \solphys, 290, 2211

\bibitem[{{Hannah} \& {Kontar}(2013)}]{Hannah}
{Hannah}, I.~G., \& {Kontar}, E.~P. 2013, \aap, 553, A10

\bibitem[{{Hayes} {et~al.}(2019){Hayes}, {Gallagher}, {Dennis}, {Ireland},
  {Inglis}, \& {Morosan}}]{Hayes}
{Hayes}, L.~A., {Gallagher}, P.~T., {Dennis}, B.~R., {et~al.} 2019, \apj, 875,
  33

\bibitem[{{Hirayama}(1974)}]{Hirayama}
{Hirayama}, T. 1974, \solphys, 34, 323

\bibitem[{{Janvier} {et~al.}(2014){Janvier}, {Aulanier}, {Bommier},
  {Schmieder}, {D{\'e}moulin}, \& {Pariat}}]{Janvier}
{Janvier}, M., {Aulanier}, G., {Bommier}, V., {et~al.} 2014, \apj, 788, 60

\bibitem[{{Janvier} {et~al.}(2013){Janvier}, {Aulanier}, {Pariat}, \&
  {D{\'e}moulin}}]{Janvier2013}
{Janvier}, M., {Aulanier}, G., {Pariat}, E., \& {D{\'e}moulin}, P. 2013, \aap,
  555, A77

\bibitem[{{Judge}(2007)}]{Judge2007}
{Judge}, P.~G. 2007, \apj, 662, 677

\bibitem[{{Judge}(2010)}]{Judge2010}
{Judge}, P.~G. 2010, \apj, 708, 1238

\bibitem[{{Judge} \& {Casini}(2001)}]{Judge+Casini2001}
{Judge}, P.~G., \& {Casini}, R. 2001, in Astronomical Society of the Pacific
  Conference Series, Vol. 236, Advanced Solar Polarimetry -- Theory,
  Observation, and Instrumentation, ed. M.~{Sigwarth}, 503

\bibitem[{{Kawate} {et~al.}(2016){Kawate}, {Keenan}, \& {Jess}}]{Kawate2016}
{Kawate}, T., {Keenan}, F.~P., \& {Jess}, D.~B. 2016, \apj, 826, 3

\bibitem[{{Kay} {et~al.}(2017){Kay}, {Gopalswamy}, {Xie}, \&
  {Yashiro}}]{Kay2017}
{Kay}, C., {Gopalswamy}, N., {Xie}, H., \& {Yashiro}, S. 2017, \solphys, 292,
  78

\bibitem[{{Kopp} \& {Pneuman}(1976)}]{Kopp}
{Kopp}, R.~A., \& {Pneuman}, G.~W. 1976, \solphys, 50, 85

\bibitem[{{Landi} {et~al.}(2016){Landi}, {Habbal}, \& {Tomczyk}}]{Landi2016}
{Landi}, E., {Habbal}, S.~R., \& {Tomczyk}, S. 2016, Journal of Geophysical
  Research (Space Physics), 121, 8237

\bibitem[{{Lemen} {et~al.}(2012){Lemen}, {Title}, {Akin}, {Boerner}, {Chou},
  {Drake}, {Duncan}, {Edwards}, {Friedlaender}, {Heyman}, {Hurlburt}, {Katz},
  {Kushner}, {Levay}, {Lindgren}, {Mathur}, {McFeaters}, {Mitchell}, {Rehse},
  {Schrijver}, {Springer}, {Stern}, {Tarbell}, {Wuelser}, {Wolfson}, {Yanari},
  {Bookbinder}, {Cheimets}, {Caldwell}, {Deluca}, {Gates}, {Golub}, {Park},
  {Podgorski}, {Bush}, {Scherrer}, {Gummin}, {Smith}, {Auker}, {Jerram},
  {Pool}, {Soufli}, {Windt}, {Beardsley}, {Clapp}, {Lang}, \&
  {Waltham}}]{Lemen}
{Lemen}, J.~R., {Title}, A.~M., {Akin}, D.~J., {et~al.} 2012, \solphys, 275, 17

\bibitem[{{Lin} {et~al.}(2002){Lin}, {Dennis}, {Hurford}, {Smith}, {Zehnder},
  {Harvey}, {Curtis}, {Pankow}, {Turin}, {Bester}, {Csillaghy}, {Lewis},
  {Madden}, {van Beek}, {Appleby}, {Raudorf}, {McTiernan}, {Ramaty}, {Schmahl},
  {Schwartz}, {Krucker}, {Abiad}, {Quinn}, {Berg}, {Hashii}, {Sterling},
  {Jackson}, {Pratt}, {Campbell}, {Malone}, {Landis}, {Barrington-Leigh},
  {Slassi-Sennou}, {Cork}, {Clark}, {Amato}, {Orwig}, {Boyle}, {Banks},
  {Shirey}, {Tolbert}, {Zarro}, {Snow}, {Thomsen}, {Henneck}, {McHedlishvili},
  {Ming}, {Fivian}, {Jordan}, {Wanner}, {Crubb}, {Preble}, {Matranga}, {Benz},
  {Hudson}, {Canfield}, {Holman}, {Crannell}, {Kosugi}, {Emslie}, {Vilmer},
  {Brown}, {Johns-Krull}, {Aschwanden}, {Metcalf}, \& {Conway}}]{Lin}
{Lin}, R.~P., {Dennis}, B.~R., {Hurford}, G.~J., {et~al.} 2002, \solphys, 210,
  3

\bibitem[{{Longcope} {et~al.}(2018){Longcope}, {Unverferth}, {Klein},
  {McCarthy}, \& {Priest}}]{Longcope}
{Longcope}, D., {Unverferth}, J., {Klein}, C., {McCarthy}, M., \& {Priest}, E.
  2018, \apj, 868, 148

\bibitem[{{Morgan} \& {Druckm{\"u}ller}(2014)}]{Morgan}
{Morgan}, H., \& {Druckm{\"u}ller}, M. 2014, \solphys, 289, 2945

\bibitem[{{Narukage} {et~al.}(2011){Narukage}, {Sakao}, {Kano}, {Hara},
  {Shimojo}, {Bando}, {Urayama}, {Deluca}, {Golub}, {Weber}, {Grigis},
  {Cirtain}, \& {Tsuneta}}]{Narukage2011}
{Narukage}, N., {Sakao}, T., {Kano}, R., {et~al.} 2011, \solphys, 269, 169

\bibitem[{{Omodei} {et~al.}(2018){Omodei}, {Pesce-Rollins}, {Longo},
  {Allafort}, \& {Krucker}}]{Omodei}
{Omodei}, N., {Pesce-Rollins}, M., {Longo}, F., {Allafort}, A., \& {Krucker},
  S. 2018, \apjl, 865, L7

\bibitem[{{Polito} {et~al.}(2018{\natexlab{a}}){Polito}, {Dud{\'\i}k},
  {Ka{\v{s}}parov{\'a}}, {Dzif{\v{c}}{\'a}kov{\'a}}, {Reeves}, {Testa}, \&
  {Chen}}]{Polito2018}
{Polito}, V., {Dud{\'\i}k}, J., {Ka{\v{s}}parov{\'a}}, J., {et~al.}
  2018{\natexlab{a}}, \apj, 864, 63

\bibitem[{{Polito} {et~al.}(2018{\natexlab{b}}){Polito}, {Galan}, {Reeves}, \&
  {Musset}}]{Polito}
{Polito}, V., {Galan}, G., {Reeves}, K.~K., \& {Musset}, S. 2018{\natexlab{b}},
  \apj, 865, 161

\bibitem[{{Polito} {et~al.}(2015){Polito}, {Reeves}, {Del Zanna}, {Golub}, \&
  {Mason}}]{Polito2015}
{Polito}, V., {Reeves}, K.~K., {Del Zanna}, G., {Golub}, L., \& {Mason}, H.~E.
  2015, \apj, 803, 84

\bibitem[{{Savage} {et~al.}(2010){Savage}, {McKenzie}, {Reeves}, {Forbes}, \&
  {Longcope}}]{Savage2010}
{Savage}, S.~L., {McKenzie}, D.~E., {Reeves}, K.~K., {Forbes}, T.~G., \&
  {Longcope}, D.~W. 2010, \apj, 722, 329

\bibitem[{{Seaton} {et~al.}(2017){Seaton}, {Bartz}, \& {Darnel}}]{Seaton2017}
{Seaton}, D.~B., {Bartz}, A.~E., \& {Darnel}, J.~M. 2017, \apj, 835, 139

\bibitem[{{Sturrock}(1968)}]{Sturrock}
{Sturrock}, P.~A. 1968, in IAU Symposium, Vol.~35, Structure and Development of
  Solar Active Regions, ed. K.~O. {Kiepenheuer}, 471

\bibitem[{{Tomczyk} {et~al.}(2008){Tomczyk}, {Card}, {Darnell}, {Elmore},
  {Lull}, {Nelson}, {Streander}, {Burkepile}, {Casini}, \& {Judge}}]{CoMP}
{Tomczyk}, S., {Card}, G.~L., {Darnell}, T., {et~al.} 2008, \solphys, 247, 411

\bibitem[{{van Driel-Gesztelyi} {et~al.}(1997){van Driel-Gesztelyi}, {Wiik},
  {Schmieder}, {Tarbell}, {Kitai}, {Funakoshi}, \& {Anwar}}]{vanDriel1997}
{van Driel-Gesztelyi}, L., {Wiik}, J.~E., {Schmieder}, B., {et~al.} 1997,
  \solphys, 174, 151

\bibitem[{{Warren} {et~al.}(2018){Warren}, {Brooks}, {Ugarte-Urra}, {Reep},
  {Crump}, \& {Doschek}}]{Warren}
{Warren}, H.~P., {Brooks}, D.~H., {Ugarte-Urra}, I., {et~al.} 2018, \apj, 854,
  122

\bibitem[{{Yu} {et~al.}(2020){Yu}, {Chen}, {Reeves}, {Gary}, {Musset},
  {Fleishman}, {Nita}, \& {Glesener}}]{Yu2020}
{Yu}, S., {Chen}, B., {Reeves}, K.~K., {et~al.} 2020, arXiv e-prints,
  arXiv:2007.10443

\end{thebibliography}

\end{document}